\documentclass[a4paper]{article}

\usepackage[english]{babel}
\usepackage[utf8]{inputenc}
\usepackage[T1]{fontenc}
\usepackage{amsmath}
\usepackage{graphicx}
\usepackage[colorinlistoftodos]{todonotes}
\usepackage[a4paper,left=2.5cm,right=2.5cm,top=2.cm,bottom=2cm]{geometry}
\usepackage{authblk}
\usepackage{dsfont}
\usepackage{txfonts}
\usepackage{bbold} 
\usepackage{amsopn}  
\usepackage{amssymb}
\usepackage{graphicx}
\newtheorem{example}{Example}

\newtheorem{definition}{Definition}

\title{Permissible extensions of classical to quantum games combining three strategies}

\author{
  Piotr Frąckiewicz\\
  \textit{Institute of Exact and Technical Sciences, Pomeranian University in Słupsk, Poland} \\
  piotr.frackiewicz@upsl.edu.pl
  \and
  Marek Szopa\\
  \textit{Department of Operations Research,
        University of Economics in Katowice, Poland} \\
        marek.szopa@uekat.pl
}

\date{\today}

\begin{document}
\maketitle

\begin{abstract}
We study the extension of classical games to the quantum domain, generated by the addition of one unitary strategy to two classical strategies of each player. The conditions that need to be met by unitary operations to ensure that the extended game is invariant with respect to the isomorphic transformations of the input game are determined. It has been shown that there are three types of these extensions, two of them are purely quantum. On the other hand, it has been demonstrated that the extensions of two versions of the same classical game by a unitary operator that does not meet these conditions may result in quantum games that are non-equivalent, e.g. having different Nash equilibria. We use the obtained results to extend the classical Prisoner's Dilemma game to a quantum game that has a unique Nash equilibrium closer to Pareto-optimal solutions than the original one.
\end{abstract}
\section{Introduction}
Quantum game theory is an interdisciplinary research area that combines game theory and quantum computation theory. Recent research has explored the potential of extensions of classical to quantum games that could offer new insights into problem solving strategies. One of the pioneering studies that demonstrated the amalgamation of game theory and quantum computing was \cite{meyer_quantum_1999}. Meyer presented the extensive-form PQ Penny Flip game, demonstrating that a player equipped with quantum strategy in the form of the Hadamard matrix has an advantage over a player using classical strategies. 
However, one of the primary objectives of quantum game theory is to expand the repertoire of classical games with new strategies available in the quantum domain. Thanks to the use of new strategies, players who were not able to achieve mutually satisfactory results when playing classically (e.g. in the Prisoner Dilemma) can often find an equilibrium closer to or equal to Pareto-optimal solutions. This was the motivation behind Eisert's et al. work \cite{eisert_quantum_1999}, which provided a scheme for playing bimatrix games $2\times 2$ and has become the standard for extending games to the quantum domain. In this approach, quantum strategies are represented by unitary operators with which players act on an entangled state. This paper also initiated the pursuit of optimal parameterization of pure quantum strategies \cite{flitney_nash_2007,elgazzar_quantum_2020}. By pure strategies, we mean unitary operations generated within a specific parameterization according to the quantization rule introduced in \cite{eisert_quantum_1999}. Players' strategies should be parameterized in such a way that, on the one hand, does not provide too much choice, which then results in a lack of nontrivial Nash equilibria, and, on the other hand, yields quantum expansion which does not affect the invariance of the original game with regard to isomorphic transformations \cite{frackiewicz_strong_2016}. In one of our earlier papers \cite{frackiewicz_nash_2022} we  investigated pure strategies Nash equilibria of \mbox{$2\times 2$~bimatrix} games in the parametrization introduced in \cite{frackiewicz_quantum_2017}. As was demonstrated in \cite{frackiewicz_nash_2022}, there are two-parameter unitary operations that are invariant under isomorphic transformations of the input classical game and, at the same time, admit for non-trivial Nash equilibria.

The second approach, which simplifies the Eisert-Wilkens-Lewenstein (EWL) scheme, was developed over time on the basis of mixed strategies. In this approach quantum games are considered as classical games extended by a finite number of unitary strategies. One of the well-known works in this field is \cite{ozdemir_samaritans_2003}. This article examines the effects of using quantum correlations and unitary strategies on the outcome of the Samaritan's Dilemma game. Another paper examines a finite number of unitary strategies in multi-player evolutionary game problems \cite{pawela_quantum_2013}. Some studies have focused on finding and classifying Pareto optimal Nash equilibria \cite{landsburg_nash_2011,szopa_efficiency_2021}. One can also find a number of works dedicated to quantum games on networks with finite strategy sets \cite{pawela_quantum_2016,yong_entanglement_2016,li_entanglement_2015,deng_novel_2016}. 

 The aim of the present paper is to identify, using the second approach, the necessary conditions for the unitary matrix of a quantum game extension to maintain invariance under isomorphic transformations of the classical game. We show that there are two types of quantum games that maintain such an invariance. The significance of our efforts in quantum game theory is to indicate how admissible extensions of classical games should be constructed in order to avoid undesirable dependence of the extension on the structure of the original classical game - the problem that occurs for example, in the case of extension by the strategy $Q$ defined in \cite{eisert_quantum_1999,van_enk_classical_2002}. This research is important in the context of quantum game theory, an interdisciplinary field that blends game theory with quantum computation theory. Quantum game theory seeks to expand the range of strategies available in classical games by incorporating quantum strategies. These new strategies, represented by unitary operators, allow players to achieve outcomes in quantum games that were not possible in classical settings. For instance, in the classical Prisoner's Dilemma, players often couldn't reach mutually satisfactory results, but the introduction of quantum strategies enables equilibria closer to Pareto-optimal solutions
 
 In Section 2, we briefly recall the definition of a strong isomorphism of classical games and provide examples of different isomorphisms to be used in later parts of the paper. Section 2 also gives a brief introduction to the EWL scheme. In section 3 we introduce the $3 \times 3$ type of bimatrix game, which is examined in this paper, as an extension of any classical $2 \times 2$ game with one arbitrary unitary strategy, the same for both players. In Section 3, we give an example of an extension of a classical game that does not satisfy the condition of invariance with respect to isomorphic transformations of the original game. It is shown there that two isomorphic variants of the same classical game, when extended, lead to nonequivalent quantum games, e.g. having different Nash equilibria. The main results of the paper are contained in Section 5, where we derive the conditions that must be satisfied by the unitary extension operator to meet the requirement of invariability. These conditions lead to three types of extensions, one of which is identical to the original classical game and the other two generate significantly different quantum games. In Section 6, we use the obtained results to extend the classical Prisoner's Dilemma game to a quantum game that has a unique Nash equilibrium closer to Pareto-optimal solutions than the original classical game. The results of the paper are summarized in Section 7.

\section{Preliminaries}
To make our article self-contained, we provide the reader with an essential introduction to game theory and quantum game theory.
\subsection{Strong isomorphism}
The notion of strong isomorphism determines the classes of games that are identical up to the numbering of players and their strategies. In our paper, we focus on the problem of selecting appropriate unitary strategies in a bimatrix game. As a result, the numbering of players does not influence our considerations. In what follows, we formulate a simplified definition of strong isomorphism in a strategic form game. First, we recall the notion of strategic form game \cite{maschler_game_2020}.
\begin{definition}
A game in strategic form is a triple $\Gamma = (N, (S_{i})_{i\in N}, (u_{i})_{i\in N})$ in which 
\begin{enumerate}
\item[(i)] $N = \{1,2, \dots, p\}$ is a finite set of players;
\item[(ii)] $S_{i}$ is the set of strategies of player $i$, for each player $i\in N$;
\item[(iii)] $u_{i}\colon S_{1}\times S_{2} \times \cdots \times S_{p} \to \mathds{R}$ is a function associating each vector of strategies $s = (s_{i})_{i\in N}$ with the payoff $u_{i}(s)$ to player $i$, for every player $i\in N$.
\end{enumerate}
\end{definition}
Two-person games in strategic form are usually called bimatrix games. The rows play the role of Player 1's strategies, and the columns are identified with Player 2's strategies. Each entry in the bimatrix is a pair of numbers that represent the payoffs to the players. If $S_{1} = \{A_{0}, A_{1}, \dots, A_{n}\}$ and $S_{2} = \{B_{0}, B_{1}, \dots, B_{m}\}$ then a strategic form game can be shown as 
\begin{equation}\label{generalbimatrix}
\bordermatrix{& B_{0} & B_{1} & \dots & B_{m} \cr
A_{0} & \Delta_{00} & \Delta_{01} & \dots & \Delta_{0m} \cr
A_{1} & \Delta_{10} & \Delta_{11} & \dots & \Delta_{1m} \cr 
\vdots & \vdots & \vdots & \ddots & \vdots \cr
A_{n} & \Delta_{n0} & \Delta_{n1} & \dots & \Delta_{nm}}, \quad \text{where}~\Delta_{ij} = (a_{ij}, b_{ij}). 
\end{equation}
The notion of Nash equilibrium is a fundamental solution concept in non-cooperative game theory. It defines a strategy profile in which no player can benefit by unilaterally deviating from their chosen strategy. 
\begin{definition}\cite{maschler_game_2020}
A strategy vector $s^* = (s^*_{1}, s^*_{2}, \dots, s^*_{p})$ is a Nash equilibrium if for each player $i\in N$ and each strategy $s_{i}\in S_{i}$ the following is satisfied:
\begin{equation}
u_{i}(s^*) \geq u_{i}(s_{i}, s^*_{-i}), 
\end{equation}
where $s^*_{-i} = (s^*_{1}, \dots, s^*_{i-1}, s^*_{i+1}, \dots, s^*_{p})$. 
\end{definition}
In the case of a two-person game, Nash equilibrium can be defined as follows:
\begin{definition}
A position $(i,j)$ in a bimatrix game (\ref{generalbimatrix}) is a Nash equilibrium if 
\begin{equation}
b_{ij} \geq a_{il} \quad \text{for all} \quad l\in \{0,1, \dots, m\}
\end{equation}
and
\begin{equation}
a_{ij} \geq a_{kj} \quad \text{for all} \quad k\in \{0,1, \dots, n\}.
\end{equation}
\end{definition}
Now, we can recall the isomorphism of a strategic-form game. The definition is prepared based on \cite{gabarro_complexity_2007} (see also \cite{nash_non-cooperative_1951,peleg_canonical_1999,sudholter_canonical_2000}). Due to the fact that we focus on considering two-player games in our work, we present a simplified version that does not take into account player numbering. 
\begin{definition}
Given $\Gamma = (N, (S_{i})_{i\in N}, (u_{i})_{i\in N})$ and $\Gamma' = (N, (S'_{i})_{i\in N}, (u'_{i})_{i\in N})$, let $(\varphi_{i})_{i\in N}$ be a collection of bijections $\varphi_{i}$ from $S_{i}$ to $S'_{i}$. A collection $(\varphi_{i})_{i\in N}$ is a strong isomorphism between $\Gamma$ and $\Gamma'$ if relation 
\begin{equation}\label{isocondition}
u_{i}\bigl((s_{1}, s_{2}, \dots, s_{p})\bigr) = u'_{i}\bigl((\varphi_{1}(s_{1}), \varphi_{2}(s_{2}), \dots, \varphi_{p}(s_{p}))\bigr)
\end{equation}
holds for each $i\in N$ and each strategy profile $(s_{1}, s_{2}, \dots, s_{p}) \in S_{1} \times S_{2} \times \cdots \times S_{p}$
\end{definition}
It is worth noting that the above definition applies to every game in a strategic form. Therefore, it naturally extends to the quantum game EWL, which is also a strategic form game defined by the set of players, the sets of strategies and the payoff functions.  
\begin{example}
Let us consider two bimatrix games 
\begin{equation}\label{2gameisomorphic}
\bordermatrix{& D & E & F 
\cr A & \Delta_{00} & \Delta_{01} & \Delta_{02}\cr 
B & \Delta_{10} & \Delta_{11} & \Delta_{12} \cr 
C & \Delta_{20} & \Delta_{21} & \Delta_{22}
} \quad \text{and} \quad \bordermatrix{& D' & E' & F' 
\cr A' & \Delta_{22} & \Delta_{21} & \Delta_{20} \cr 
B' & \Delta_{02} & \Delta_{01} & \Delta_{00} \cr 
C' & \Delta_{12} & \Delta_{11} & \Delta_{10}
}.
\end{equation}
Then games (\ref{2gameisomorphic}) are isomorphic as there is a pair of bijections $(\varphi_{1}, \varphi_{2})$, 
\begin{align}
&\varphi_{1}\colon \{A,B,C\} \to \{A', B', C'\}, \quad \varphi_{1} = (A \to B', B \to C', C \to A'), \\ 
&\varphi_{2}\colon \{D,E,F\} \to \{D', E', F'\}, \quad \varphi_{2} = (D \to F', E \to E', F \to D')
\end{align}
that satisfy (\ref{isocondition}). For example, consider the strategy profile $(B,F)$. Then
\begin{equation}
u'_{1}(\varphi_{1}(B), \varphi_{2}(F)) = u'_{1}(C', D') = a_{12} = u_{1}(B,F).
\end{equation}
\end{example}
\begin{example}
Let us consider a $2\times 2$ bimatrix game
\begin{equation}
\label{bimatrixiso}
\bordermatrix{ & C & D\cr
A & \Delta_{00} & \Delta_{01} \cr
B & \Delta_{10} & \Delta_{11}
}.
\end{equation}
There are three different bimatrix games 
\begin{equation}
\label{bimatrix2}
\bordermatrix{ & C' & D'\cr
A' & \Delta'_{00} & \Delta'_{01} \cr
B' & \Delta'_{10} & \Delta'_{11}
}.
\end{equation}
that are isomorphic with (\ref{bimatrixiso}). They are defined by three bijections that satisfy (\ref{isocondition}).
\begin{enumerate}
    \item $\varphi_{1} = (A \to B', B \to A'), \quad \varphi_{2} = (C \to C', D \to D')$. Then
    \begin{equation}
    \begin{aligned}
&u_{i}(A,C) = u'_{i}(B',C') \Leftrightarrow \Delta_{00} = \Delta'_{10}, \\
&u_{i}(A,D) = u'_{i}(B',D') \Leftrightarrow \Delta_{01} = \Delta'_{11}, \\
&u_{i}(B,C) = u'_{i}(A',C') \Leftrightarrow \Delta_{10} = \Delta'_{00}, \\
&u_{i}(B,D) = u'_{i}(A',D') \Leftrightarrow \Delta_{11} = \Delta'_{01}.
\end{aligned}
    \end{equation}
and bimatrix game (\ref{bimatrix2}) takes the form 
\begin{equation}
\label{bimatrixwiersz}
\bordermatrix{ & C' & D'\cr
A' & \Delta_{10} & \Delta_{11} \cr
B' & \Delta_{00} & \Delta_{01}
} \quad \text{game (\ref{bimatrixiso}) with swapped rows.}
\end{equation}
\item $\varphi_{1} = (A \to A', B \to B'), \quad \varphi_{2} = (C \to D', D \to C')$. Then
\begin{equation}
    \begin{aligned}
&u_{i}(A,C) = u'_{i}(A',D') \Leftrightarrow \Delta_{00} = \Delta'_{01}, \\
&u_{i}(A,D) = u'_{i}(A',C') \Leftrightarrow \Delta_{01} = \Delta'_{00}, \\
&u_{i}(B,C) = u'_{i}(B',D') \Leftrightarrow \Delta_{10} = \Delta'_{11}, \\
&u_{i}(B,D) = u'_{i}(B',C') \Leftrightarrow \Delta_{11} = \Delta'_{10}.
\end{aligned}
    \end{equation}
and bimatrix game (\ref{bimatrix2}) takes the form
\begin{equation}
\label{bimatrixkolumna}
\bordermatrix{ & C' & D'\cr
A' & \Delta_{01} & \Delta_{00} \cr
B' & \Delta_{11} & \Delta_{10}
} \quad \text{game (\ref{bimatrixiso}) with swapped columns.}
\end{equation}
\item $\varphi_{1} = (A \to B', B \to A'), \quad \varphi_{2} = (C \to D', D \to C')$. Then
\begin{equation}
    \begin{aligned}
&u_{i}(A,C) = u'_{i}(B',D') \Leftrightarrow \Delta_{00} = \Delta'_{11}, \\
&u_{i}(A,D) = u'_{i}(B',C') \Leftrightarrow \Delta_{01} = \Delta'_{10}, \\
&u_{i}(B,C) = u'_{i}(A',D') \Leftrightarrow \Delta_{10}  = \Delta'_{01}, \\
&u_{i}(B,D) = u'_{i}(A',C') \Leftrightarrow \Delta_{11} = \Delta'_{00}.
\end{aligned}
    \end{equation}
and bimatrix game (\ref{bimatrix2}) takes the form
\begin{equation}
\label{bimatrixwierszkolumna}
\bordermatrix{ & C' & D'\cr
A' & \Delta_{11} & \Delta_{10} \cr
B' & \Delta_{01} & \Delta_{00}
} \quad \text{game (\ref{bimatrixiso}) with swapped rows and columns.}
\end{equation}
\end{enumerate}
\end{example}
\subsection{Eisert-Wilkens-Lewenstein scheme}
In this section, we recall how to calculate payoffs according to the EWL-type scheme for $2\times 2$ games (\ref{bimatrixiso}).
Unitary operators from $\mathsf{SU}(2)$ play the role of strategies in the EWL model. The commonly used parametrization of the unitary strategy is 
\begin{equation}
U(\theta, \alpha, \beta) = \begin{pmatrix}
e^{i\alpha}\cos\frac{\theta}{2} & ie^{i\beta}\sin\frac{\theta}{2} \\ ie^{-i\beta}\sin\frac{\theta}{2} & e^{-i\alpha}\cos\frac{\theta}{2}
\end{pmatrix}, \quad \theta\in [0,\pi], \alpha, \beta \in [0,2\pi).
\end{equation}
The players by choosing $U_{1}(\theta_{1}, \alpha_{1}, \beta_{1})$ and $U_{2}(\theta_{2}, \alpha_{2}, \beta_{2})$ set the final state of the form 
\begin{equation}
|\Psi\rangle = J^{\dag}\left(U_{1}(\theta_{1}, \alpha_{1}, \beta_{1})\otimes U_{2}(\theta_{2}, \alpha_{2}, \beta_{2}) \right)J|00\rangle, 
\end{equation}
where $J = (I\otimes I + i\sigma_{x}\otimes \sigma_{x})/\sqrt{2}$ is the full entanglement operator, which case is considered throughout the paper.

Player $i$'s payoff $u_{i}$ corresponding to a strategy profile $\left(U_{1}(\theta_{1}, \alpha_{1}, \beta_{1}), U_{2}(\theta_{2}, \alpha_{2}, \beta_{2})\right)$ is defined as the average value of measurement $M_{i}$,
\begin{equation}
    M_{1} = \sum_{i,j\in {0,1}} a_{ij}|ij\rangle \langle ij|, \quad M_{2} = \sum_{i,j\in {0,1}} b_{ij}|ij\rangle \langle ij| 
\end{equation}
according to the formula
\begin{equation}\label{payoffM}
u_{i}(U_{1}(\theta_{1}, \alpha_{1}, \beta_{1}), U_{2}(\theta_{2}, \alpha_{2}, \beta_{2})) = \langle \Psi|M_{i}|\Psi\rangle.
\end{equation}
For convienience, we derive an explicit formula for (\ref{payoffM}),
\begin{align}\label{generalEWLpayoff}
    &u_{i}(U_{1}(\theta_{1}, \alpha_{1}, \beta_{1}), U_{2}(\theta_{2}, \alpha_{2}, \beta_{2})) \nonumber\\ &\quad = c^{i}_{00}\left(\cos(\alpha_{1} + \alpha_{2})\cos\frac{\theta_{1}}{2}\cos\frac{\theta_{2}}{2} + \sin(\beta_{1} + \beta_{2})\sin\frac{\theta_{1}}{2}\sin\frac{\theta_{2}}{2}\right)^2 \nonumber\\
    &\quad \quad + c^i_{01}\left(\cos(\alpha_{1} - \beta_{2})\cos\frac{\theta_{1}}{2}\sin\frac{\theta_{2}}{2} + \sin(\alpha_{2} - \beta_{1})\sin\frac{\theta_{1}}{2}\cos\frac{\theta_{2}}{2}\right)^2 \nonumber \\ 
    &\quad \quad + c^i_{10}\left(\sin(\alpha_{1} - \beta_{2})\cos\frac{\theta_{1}}{2}\sin\frac{\theta_{2}}{2} + \cos(\alpha_{2} - \beta_{1})\sin\frac{\theta_{1}}{2}\cos\frac{\theta_{2}}{2}\right)^2 \nonumber \\ 
    &\quad \quad + c^i_{11}\left(\sin(\alpha_{1} + \alpha_{2})\cos\frac{\theta_{1}}{2}\cos\frac{\theta_{2}}{2} - \cos(\beta_{1} + \beta_{2})\sin\frac{\theta_{1}}{2}\sin\frac{\theta_{2}}{2}\right)^2
\end{align}
where $c^1_{ij} = a_{ij}$ and $c^2_{ij} = b_{ij}$ are the payoffs of (\ref{bimatrixiso}).
\section{A classical game with an additional unitary strategy} \label{section3}
In this section, we will consider a game with two classical strategies and one additional strategy, which is given by any unitary operator the same for both players. Such a game can be realized using the EWL model, in which the first and second classical strategy can be identified with the identity operator $I$ and $iX$, respectively. Given (\ref{bimatrixiso}) and substituting $I = U(0,0,0)$, $iX = U(\pi, 0,0)$ and a general unitary strategy $U(\theta, \alpha, \beta)$ to formula (\ref{generalEWLpayoff}) we obtain the following $3\times 3$ bimatrix game:
\begin{equation}\label{quantum1}
\bordermatrix{ & I & iX & U \cr I & \Delta_{00} & \Delta_{01} & u(I, U) \cr 
iX & \Delta_{10} & \Delta_{11} & u(iX, U) \cr 
U & u(U, I) & u(U, iX) & u(U, U)
}, \quad \text{where}
\end{equation}
\begin{align}
&u(U, I) = \Delta_{00}\cos^2\alpha\cos^2\frac{\theta}{2} + \Delta_{01}\sin^2\beta\sin^2\frac{\theta}{2} + \Delta_{10}\cos^2\beta\sin^2\frac{\theta}{2} + \Delta_{11}\sin^2\alpha\cos^2\frac{\theta}{2},\\
&u(U, iX) =  \Delta_{00}\sin^2\beta\sin^2\frac{\theta}{2} + \Delta_{01}\cos^2\alpha\cos^2\frac{\theta}{2} + \Delta_{10}\sin^2\alpha\cos^2\frac{\theta}{2} + \Delta_{11}\cos^2\beta\sin^2\frac{\theta}{2},\\
&u(I, U) = \Delta_{00}\cos^2\alpha\cos^2\frac{\theta}{2} + \Delta_{01}\cos^2\beta\sin^2\frac{\theta}{2} + \Delta_{10}\sin^2\beta\sin^2\frac{\theta}{2} + \Delta_{11}\sin^2\alpha\cos^2\frac{\theta}{2},\\
&u(iX, U) = \Delta_{00}\sin^2\beta\sin^2\frac{\theta}{2} + \Delta_{01}\sin^2\alpha\cos^2\frac{\theta}{2} + \Delta_{10}\cos^2\alpha\cos^2\frac{\theta}{2} + \Delta_{11}\cos^2\beta\sin^2\frac{\theta}{2},
\end{align}
\begin{align}\label{wyplatauu}
u(U, U) &= \Delta_{00}\left(\cos(2\alpha)\cos^2\frac{\theta}{2} + \sin(2\beta)\sin^2\frac{\theta}{2}\right)^2 \nonumber \\ 
&\quad + 
\frac{1}{4}(\Delta_{01}+\Delta_{10})\left(\cos(\alpha-\beta) + \sin(\alpha-\beta)\right)^2\sin^2\theta \nonumber \\
&\quad +
\Delta_{11}\left(\sin(2\alpha)\cos^2\frac{\theta}{2} - \cos(2\beta)\sin^2\frac{\theta}{2}\right)^2.
\end{align}
\section{Significance of invariance with respect to isomorphic forms of the initial game}
Let us consider the game (\ref{bimatrixiso}) and its isomorphic counterpart (\ref{bimatrixwiersz}) resulting from reversing the first player's strategy. Let us then consider as an additional strategy the well-known operator $Q = U(0, \pi/2, 0)$, that is meant to solve the Prisoner's Dilemma game \cite{eisert_quantum_1999,van_enk_classical_2002}. In that case, according to the scheme (\ref{quantum1})-(\ref{wyplatauu}),
\begin{equation}\label{strz1}
\bordermatrix{ & C & D\cr
A & \Delta_{00} & \Delta_{01} \cr
B & \Delta_{10} & \Delta_{11}
} \xrightarrow[]{\text{EWL model}} \bordermatrix{& I & iX & Q \cr
    I & \Delta_{00} & \Delta_{01} & \Delta_{11} \cr
    iX & \Delta_{10} & \Delta_{11} & \Delta_{01} \cr 
    Q & \Delta_{11} & \Delta_{10} & \Delta_{00}}.
\end{equation}
On the other hand, a game with the first player's strategies reversed implies the following $3\times 3$ game:
\begin{equation}\label{strz2}
        \bordermatrix{ & C & D\cr
B & \Delta_{10} & \Delta_{11} \cr
A & \Delta_{00} & \Delta_{01}
} \xrightarrow[]{\text{EWL model}}     \bordermatrix{& I & iX & Q \cr
    I & \Delta_{10} & \Delta_{11} & \Delta_{01} \cr
    iX & \Delta_{00} & \Delta_{01} & \Delta_{11} \cr 
    Q & \Delta_{01} & \Delta_{00} & \Delta_{10}}.
    \end{equation}
It is obvious that $2\times 2$ bimatrices in (\ref{strz1}) and (\ref{strz2}) represent the same problem with respect to a final result of a game, in particular, in terms of payoffs predicted by Nash equilibrium. In contrast, games with additional unitary strategy $Q$ are radically different. To illustrate this, let's consider a specific example of the Prisoner's Dilemma game
\begin{equation}\label{pdkonkret}
\bordermatrix{ & C & D \cr
C & (3,3) & (0,5) \cr
D & (5,0) & (1,1)}.
\end{equation}
The game has a unique Nash equilibrium with the payoff outcome $(1,1)$, and swapping the order of player 1's strategies has no influence on that outcome. If we compare the $3\times 3$ games of (\ref{strz1}) and (\ref{strz2}), 
\begin{equation}\label{PD3a}
\bordermatrix{& I & iX & Q \cr
    I & (3,3) & (0,5) & (1,1) \cr
    iX & (5,0) & (1,1) & (0,5) \cr 
    Q & (1,1) & (5,0) & (3,3)}, \quad 
    \bordermatrix{& I & iX & Q \cr
    I & (5,0) & (1,1) & (0,5) \cr
    iX & (3,3) & (0,5) & (1,1) \cr 
    Q & (0,5) & (3,3) & (5,0)},
\end{equation}
the first one has a unique Nash equilibrium $(Q,Q)$ (with payoff outcome $(3,3)$) whereas the second one results with the mixed Nash equilibrium $\bigl((1/2, 0, 1/2), (1/2, 0, 1/2)\bigr)$ (with payoff outcome $(5/2, 5/2)$). 

The same applies when considering $2\times 2$ games that differ in the order of columns. In that case we obtain a game
\begin{equation}\label{PD3b}
\bordermatrix{& I & iX & Q \cr
    I & (0,5) & (3,3) & (5,0) \cr
    iX & (1,1) & (5,0) & (3,3) \cr 
    Q & (5,0) & (1,1) & (0,5)}
\end{equation}
with a Nash equilibrium $\bigl((1/2, 0, 1/2), (1/2, 0, 1/2)\bigr)$. 

We will obtain yet another result if we consider the Prisoner's Dilemma game in which the order of both rows and columns is switched. In general case, 
\begin{equation}
        \begin{pmatrix}
            \Delta_{11} & \Delta_{10} \\
            \Delta_{01} & \Delta_{00} 
        \end{pmatrix} \xrightarrow[]{\text{EWL model}}     \bordermatrix{& I & iX & Q \cr
    I & \Delta_{11} & \Delta_{10} & \Delta_{00} \cr
    iX & \Delta_{01} & \Delta_{00} & \Delta_{10} \cr 
    Q & \Delta_{00} & \Delta_{01} & \Delta_{11}}.
    \end{equation} 
By substituting payoffs from (\ref{pdkonkret}) we will obtain a $3\times 3$ game
\begin{equation}\label{gi_3}
    \bordermatrix{& I & iX & Q \cr
    I & (1,1) & (5,0) & (3,3) \cr
    iX & (0,5) & (3,3) & (5,0) \cr 
    Q & (3,3) & (0,5) & (1,1)}
\end{equation}
with a unique Nash equilibrium $\bigl((14/25, 2/25, 9/25), (14/25, 2/25, 9/25)\bigr)$ determining payoff 51/25 for each player.

We observe that the extension of the classical game by one unitary operator (\ref{quantum1}) can lead to nonequivalent games depending on the representation of the classical game. As we showed above, the extension by the $Q$ operator is sensitive to the reordering of the classical game strategy. The games (\ref{PD3a}), (\ref{PD3b}) and (\ref{gi_3}) obtained by the addition of a new strategy $Q$, which violates the invariance to isomorphic transformation of the initial game, are in fact new games which do not solve the original Prisoner's Dilemma. Such an observation, albeit with different reasoning, was made in relation to the first game of (\ref{PD3a}) by Enk and Pike \cite{van_enk_classical_2002}. The problem can be solved by appropriately selecting the operator $U$. In the next section, we will determine the class of operators that extend the game in an unambiguous way.
\section{Derivation of conditions for invariance under isomorphic transformations of the game}
In this section, we derive equations that must be satisfied by unitary operations in (\ref{quantum1}) to ensure the invariance of the game with respect to isomorphic transformations of the classical game (\ref{bimatrixiso}). First, consider an isomorphic counterpart of the game (\ref{bimatrixiso}) with swapped rows, that is, a game in the form of (\ref{bimatrixwiersz}). According to the protocol described in Section~\ref{section3}, the EWL model corresponding to (\ref{bimatrixwiersz}) implies the following $3\times 3$ game:
\begin{equation}\label{wbimacierz}
\bordermatrix{ & I & iX & U \cr I & \Delta_{10} & \Delta_{11} & u^r(I, U) \cr 
iX & \Delta_{00} & \Delta_{01} & u^r(iX, U) \cr 
U & u^r(U, I) & u^r(U, iX) & u^r(U, U)
}, \quad \text{where}
\end{equation}
\begin{align}
&u^r(U, I) = \Delta_{10}\cos^2\alpha\cos^2\frac{\theta}{2} + \Delta_{11}\sin^2\beta\sin^2\frac{\theta}{2} + \Delta_{00}\cos^2\beta\sin^2\frac{\theta}{2} + \Delta_{01}\sin^2\alpha\cos^2\frac{\theta}{2},\\
&u^r(U, iX) =  \Delta_{10}\sin^2\beta\sin^2\frac{\theta}{2} + \Delta_{11}\cos^2\alpha\cos^2\frac{\theta}{2} + \Delta_{00}\sin^2\alpha\cos^2\frac{\theta}{2} + \Delta_{01}\cos^2\beta\sin^2\frac{\theta}{2},\\
&u^r(I, U) = \Delta_{10}\cos^2\alpha\cos^2\frac{\theta}{2} + \Delta_{11}\cos^2\beta\sin^2\frac{\theta}{2} + \Delta_{00}\sin^2\beta\sin^2\frac{\theta}{2} + \Delta_{01}\sin^2\alpha\cos^2\frac{\theta}{2},\\
&u^r(iX, U) = \Delta_{10}\sin^2\beta\sin^2\frac{\theta}{2} + \Delta_{11}\sin^2\alpha\cos^2\frac{\theta}{2} + \Delta_{00}\cos^2\alpha\cos^2\frac{\theta}{2} + \Delta_{01}\cos^2\beta\sin^2\frac{\theta}{2},
\end{align}
\begin{align}
u^r(U_{1}, U_{2}) &= \Delta_{10}\left(\cos(2\alpha)\cos^2\frac{\theta}{2} + \sin(2\beta)\sin^2\frac{\theta}{2}\right)^2\\ 
&\quad + 
\frac{1}{4}(\Delta_{00} + \Delta_{11})\left(\cos(\alpha-\beta) + \sin(\alpha-\beta)\right)^2\sin^2\theta \\
&\quad +
\Delta_{01}\left(\sin(2\alpha)\cos^2\frac{\theta}{2} - \cos(2\beta)\sin^2\frac{\theta}{2}\right)^2.
\end{align}
Bimatrix game (\ref{bimatrixiso}) is assumed to be generic, thus comparing (\ref{quantum1}) and (\ref{wbimacierz}) we see that
\begin{equation}\label{klasycznywarunek1}
u(I,I) = u'(iX, I), ~u(I,iX) = u'(iX, iX), ~u(iX,I) = u'(I, I), ~u(iX,iX) = u'(I, iX).
\end{equation}
Taking into account all possible bijections $(\varphi_{1}, \varphi_{2})$, we have shown that the only pair for which the equality of (\ref{quantum1}) and (\ref{wbimacierz}) can hold is
\begin{equation}\label{bijectionwiersz}
\varphi_{1} = (I \to iX, iX \to I, U \to U), \quad \varphi_{2} = (I \to I, iX \to iX, U \to U). 
\end{equation}
Considering (\ref{klasycznywarunek1}), mappings (\ref{bijectionwiersz}) satisfy condition (\ref{isocondition}) if
\begin{align}
&u(I,U) = u'(\varphi_{1}(I), \varphi_{2}(U)) = u^r(iX,U), \label{weq0}\\
&u(iX,U) = u'(\varphi_{1}(iX), \varphi_{2}(U)) = u^r(I, U),\label{weq00}\\
&u(U, I) = u'(\varphi_{1}(U), \varphi_{2}(I)) = u^r(U,I), \label{weq1}\\ 
&u(U, iX) = u'(\varphi_{1}(U), \varphi_{2}(iX)) = u^r(U,iX), \label{weq2}\\ 
&u(U, U) = u'(\varphi_{1}(U), \varphi_{2}(U)) = u^r(U, U).\label{weq3}
\end{align}
Equations (\ref{weq0}) and  (\ref{weq00}) are satisfied trivially. From (\ref{weq1}) and (\ref{weq2}) it follows that
\begin{align}\label{uui}
&u(U,I) - u^{r}(U,I) \nonumber \\ &\quad = \Delta_{00}\left(\cos^2\alpha\cos^2\frac{\theta}{2}-\cos^2\beta\sin^2\frac{\theta}{2} \right) + \Delta_{01}\left(\sin^2\beta\sin^2\frac{\theta}{2}-\sin^2\alpha\cos^2\frac{\theta}{2}\right) \nonumber \\ 
&\quad \quad + \Delta_{10}\left(\cos^2\beta\sin^2\frac{\theta}{2}-\cos^2\alpha\cos^2\frac{\theta}{2}\right) + \Delta_{11}\left(\sin^2\alpha\cos^2\frac{\theta}{2}-\sin^2\beta\sin^2\frac{\theta}{2} \right) = 0
\end{align}
and 
\begin{align}\label{uuix}
&u(U,iX) - u^{r}(U,iX) \nonumber \\ &\quad = \Delta_{00}\left(\sin^2\beta\sin^2\frac{\theta}{2}-\sin^2\alpha\cos^2\frac{\theta}{2} \right) + \Delta_{01}\left(\cos^2\alpha\cos^2\frac{\theta}{2}-\cos^2\beta\sin^2\frac{\theta}{2}\right) \nonumber \\ 
&\quad \quad + \Delta_{10}\left(\sin^2\alpha\cos^2\frac{\theta}{2}-\sin^2\beta\sin^2\frac{\theta}{2}\right) + \Delta_{11}\left(\cos^2\beta\sin^2\frac{\theta}{2}-\cos^2\alpha\cos^2\frac{\theta}{2} \right) = 0.
\end{align}
Equations (\ref{uui}) and (\ref{uuix}) imply the following system of equations:
\begin{equation}
\begin{cases}\label{systemwiersz}
\cos^2\alpha\cos^2\frac{\theta}{2} = \cos^2\beta\sin^2\frac{\theta}{2},\\
\sin^2\alpha\cos^2\frac{\theta}{2} = \sin^2\beta\sin^2\frac{\theta}{2}.
\end{cases}
\end{equation}
By adding both sides, we see that
$\cos^2\frac{\theta}{2} = \sin^2\frac{\theta}{2}$. Taking into account $\theta \in [0,\pi]$, we obtain $\theta = \pi/2$. As a result, (\ref{systemwiersz}) is equivalent to
\begin{equation}\label{uklad1}
\begin{cases}
\theta = \frac{\pi}{2}, \\ 
\sin^2\alpha = \sin^2\beta.
\end{cases}
\end{equation}
Bearing in mind that $\theta = \pi/2$, we can write equation (\ref{weq3}) in the form 
\begin{align}\label{weq3.1}
&u(U, U) - u^r(U, U) \nonumber \\
&\quad = \Delta_{00}\left(\left(\cos(2\alpha) + \sin(2\beta)\right)^2 -\left(\sin(\alpha-\beta) + \cos(\alpha - \beta)\right)^2  \right) \nonumber\\
&\quad \quad + \Delta_{01}\left(\left(\cos(\alpha-\beta) + \sin(\alpha - \beta)\right)^2 -\left(\sin(2\alpha) - \cos(2\beta)\right)^2  \right) \nonumber \\
&\quad \quad +  \Delta_{10}\left(\left(\sin(\alpha-\beta) + \cos(\alpha - \beta)\right)^2 -\left(\cos(2\alpha) + \sin(2\beta)\right)^2  \right) \nonumber \\
&\quad \quad + \Delta_{11}\left(\left(\sin(\alpha_{1}+\alpha_{2}) - \cos(\beta_{1} + \beta_{2})\right)^2 -\left(\cos(\alpha_{1}-\beta_{2}) + \sin(\alpha_{2} - \beta_{1})\right)^2  \right) = 0.
\end{align}
From equation (\ref{weq3.1}) it follows
\begin{equation}\label{onetwo}
\begin{cases}
\left(\cos(2\alpha) + \sin(2\beta)\right)^2 = 1 + \sin(2(\alpha - \beta)),\\
\left(\sin(2\alpha) - \cos(2\beta)\right)^2= 1 + \sin(2(\alpha - \beta)).
\end{cases}
\end{equation}
Let us apply a similar reasoning to the EWL scheme for an isomorphic counterpart of game (\ref{bimatrixiso}) with swapped columns (that is, game (\ref{bimatrixkolumna}). We show that such a procedure will not yield new equations. We have
\begin{equation}\label{kquantum}
\bordermatrix{ & I & iX & U \cr I & \Delta_{01} & \Delta_{00} & u^c(I, U) \cr 
iX & \Delta_{11} & \Delta_{10} & u^c(iX, U) \cr 
U & u^c(U, I) & u^c(U, iX) & u^c(U, U)}, \quad \text{where}
\end{equation}
\begin{align}
&u^c(U, I) = \Delta_{01}\cos^2\alpha\cos^2\frac{\theta}{2} + \Delta_{00}\sin^2\beta\sin^2\frac{\theta}{2} + \Delta_{11}\cos^2\beta\sin^2\frac{\theta}{2} + \Delta_{10}\sin^2\alpha\cos^2\frac{\theta}{2}, \\
&u^c(U, iX) = \Delta_{01}\sin^2\beta\sin^2\frac{\theta}{2} + \Delta_{00}\cos^2\alpha\cos^2\frac{\theta}{2} + \Delta_{11}\sin^2\alpha\cos^2\frac{\theta}{2} + \Delta_{10}\cos^2\beta\sin^2\frac{\theta}{2}, \\ 
&u^c(I, U) = \Delta_{01}\cos^2\alpha\cos^2\frac{\theta}{2} + \Delta_{00}\cos^2\beta\sin^2\frac{\theta}{2} + \Delta_{11}\sin^2\beta\sin^2\frac{\theta}{2} + \Delta_{10}\sin^2\alpha\cos^2\frac{\theta}{2}, \\
&u^c(iX, U) = \Delta_{01}\sin^2\beta\sin^2\frac{\theta}{2} + \Delta_{00}\sin^2\alpha\cos^2\frac{\theta}{2} + \Delta_{11}\cos^2\alpha\cos^2\frac{\theta}{2} + \Delta_{10}\cos^2\beta\sin^2\frac{\theta}{2},
\end{align}
\begin{align}
u^c(U, U) &= \Delta_{01}\left(\cos(2\alpha)\cos^2\frac{\theta}{2} + \sin(2\beta)\sin^2\frac{\theta}{2}\right)^2\\ 
&\quad + 
\frac{1}{4}(\Delta_{00} + \Delta_{11})\left(\cos(\alpha-\beta) + \sin(\alpha-\beta)\right)^2\sin^2\theta \\
&\quad +
\Delta_{10}\left(\sin(2\alpha)\cos^2\frac{\theta}{2} - \cos(2\beta)\sin^2\frac{\theta}{2}\right)^2.
\end{align}
Similarly to the previous case, we come to the conclusion that for the games (\ref{quantum1}) and (\ref{kquantum}) to be isomorphic, the bijecions $(\varphi_{1}, \varphi_{2})$ must have the form
\begin{equation}\label{bijectionkolumna}
\varphi_{1} = (I \to I, iX \to iX, U \to U), \quad \varphi_{2} = (I \to iX, iX \to I, U \to U) 
\end{equation}
and the following equations must hold
\begin{align}
&u(U,iX) = u'(\varphi_{1}(U), \varphi_{2}(iX)) = u^c(U,I), \label{keq0}\\
&u(U,I) = u'(\varphi_{1}(U), \varphi_{2}(I)) = u^c(U, iX),\label{keq00}\\
&u(I, U) = u'(\varphi_{1}(I), \varphi_{2}(U)) = u^c(I,U), \label{keq1}\\ 
&u(iX, U) = u'(\varphi_{1}(iX), \varphi_{2}(U)) = u^c(iX, U), \label{keq2}\\ 
&u(U, U) = u'(\varphi_{1}(U), \varphi_{2}(U)) = u^c(U, U)\label{keq3}.
\end{align}
In this case, equations (\ref{keq0}) and (\ref{keq00}) are trivial. From (\ref{keq1}) and (\ref{keq2}) we see that
\begin{align}
&u(I,U) - u^{c}(I,U) \nonumber \\ &\quad = \Delta_{00}\left(\cos^2\alpha\cos^2\frac{\theta}{2}-\cos^2\beta\sin^2\frac{\theta}{2} \right) + \Delta_{01}\left(\cos^2\beta\sin^2\frac{\theta}{2}-\cos^2\alpha\cos^2\frac{\theta}{2}\right) \nonumber \\ 
&\quad \quad + \Delta_{10}\left(\sin^2\beta\sin^2\frac{\theta}{2}-\sin^2\alpha\cos^2\frac{\theta}{2}\right) + \Delta_{11}\left(\sin^2\alpha\cos^2\frac{\theta}{2}-\sin^2\beta\sin^2\frac{\theta}{2} \right) = 0
\end{align}
and
\begin{align}
&u(iX,U) - u^{c}(iX,U) \nonumber \\ &\quad = 
\Delta_{00}\left(\sin^2\beta\sin^2\frac{\theta}{2}-\sin^2\alpha\cos^2\frac{\theta}{2} \right) + \Delta_{01}\left(\sin^2\alpha\cos^2\frac{\theta}{2}-\sin^2\beta\sin^2\frac{\theta}{2}\right) \nonumber \\ 
&\quad \quad + \Delta_{10}\left(\cos^2\alpha\cos^2\frac{\theta}{2}-\cos^2\beta\sin^2\frac{\theta}{2}\right) + \Delta_{11}\left(\cos^2\beta\sin^2\frac{\theta}{2}-\cos^2\alpha\cos^2\frac{\theta}{2} \right) = 0.
\end{align}
Reasoning similarly to the swapped rows case, one can see that (\ref{keq1}) and (\ref{keq2}) are equivalent to (\ref{uklad1}).
Eq. (\ref{keq3}) can be rewritten as  
\begin{align}\label{keq3.1}
&u(U, U) - u^c(U, U) \nonumber \\
&\quad = \Delta_{00}\left(\left(\cos(2\alpha) + \sin(2\beta)\right)^2 -\left(\cos(\alpha-\beta) + \sin(\alpha - \beta)\right)^2  \right) \nonumber\\
&\quad \quad + \Delta_{01}\left(\left(\cos(\alpha-\beta) + \sin(\alpha - \beta)\right)^2 -\left(\cos(2\alpha) + \sin(2\beta)\right)^2  \right) \nonumber \\
&\quad \quad +  \Delta_{10}\left(\left(\sin(\alpha-\beta) + \cos(\alpha - \beta)\right)^2 -\left(\sin(2\alpha) - \cos(2\beta)\right)^2  \right) \nonumber \\
&\quad \quad + \Delta_{11}\left(\left(\sin(2\alpha) - \cos(2\beta)\right)^2 -\left(\sin(\alpha-\beta) + \cos(\alpha - \beta)\right)^2  \right) = 0
\end{align}
what again leads to (\ref{onetwo}).
We can show that considering games with swapped rows and columns (\ref{bimatrixwierszkolumna}) will not give any new equations. As a result, the parameters $\theta$, $\alpha$ and $\beta$ are completely determined by (\ref{uklad1}) and (\ref{onetwo}).
For the solution of the system of equations Eqs (\ref{onetwo}), let's add it side by side to get
\begin{equation}\label{threefoursol}
2 - 2\sin(2(\alpha - \beta)) = 2 + 2\sin(2(\alpha - \beta))
\end{equation}
or, equivalently $\sin(2(\alpha - \beta)) = 0$, which means that
\begin{equation}\label{threefoursolution0}
\alpha - \beta = k \frac{\pi}{2}.
\end{equation}
Subtracting the equations of the system (\ref{onetwo}) from each other, we obtain the dependence
\begin{equation}\label{threefoursolution1}
\alpha + \beta = l \frac{\pi}{2}.
\end{equation}
The values $k$ and $l$ in equations (\ref{threefoursolution0}) and (\ref{threefoursolution1}) are arbitrary integers, hence $\alpha$ and $\beta$ can be determined
\begin{equation}\label{alfabeta}
\begin{cases}
\alpha =(l+k)\frac{\pi}{4}, \\
\beta =(l-k)\frac{\pi}{4} .
\end{cases}
\end{equation}
It turns out that the remaining condition $\sin^2\alpha = \sin^2\beta$ of (\ref{uklad1}) is satisfied by $\alpha$ and $\beta$ depending on the parity of $k$ and $l$. It is met if at least one of them is even, however, not if both $k$ and $l$ are odd. Therefore, taking into account $\alpha, \beta \in [0,2\pi)$, we get 24 solutions. They generate three types of $3\times 3$ bimatrix games:
\begin{enumerate}
\item[i.] $(\alpha, \beta) \in \{0,\pi\}\times \{0,\pi\}$ \hspace{5pt} ($k$, $l$ even)
\begin{equation}\label{typeI}
\bordermatrix{ & I & iX & U \cr I & \Delta_{00} & \Delta_{01} & \frac{\Delta_{00}+\Delta_{01}}{2} \cr 
iX & \Delta_{10} & \Delta_{11} & \frac{\Delta_{10}+\Delta_{11}}{2} \cr 
U & \frac{\Delta_{00}+\Delta_{10}}{2} & \frac{\Delta_{01}+\Delta_{11}}{2} & \frac{\Delta_{00}+\Delta_{01} + \Delta_{10} + \Delta_{11}}{4}
}, 
\end{equation}
\item[ii.] $(\alpha, \beta) \in \{\pi/2,3\pi/2\}\times \{\pi/2,3\pi/2\}$ \hspace{5pt} ($k$, $l$ even)
\begin{equation}\label{typeII}
\bordermatrix{ & I & iX & U \cr I & \Delta_{00} & \Delta_{01} & \frac{\Delta_{10}+\Delta_{11}}{2} \cr 
iX & \Delta_{10} & \Delta_{11} & \frac{\Delta_{00}+\Delta_{01}}{2} \cr 
U & \frac{\Delta_{01}+\Delta_{11}}{2} & \frac{\Delta_{00}+\Delta_{10}}{2} & \frac{\Delta_{00}+\Delta_{01} + \Delta_{10} + \Delta_{11}}{4}
}, 
\end{equation}
\item[iii.] $(\alpha, \beta) \in \{\pi/4,3\pi/4, 5\pi/4, 7\pi/4\}\times \{\pi/4,3\pi/4, 5\pi/4, 7\pi/4\}$ \hspace{5pt} ($k$, $l$ of different parity)
\begin{equation}\label{typeIII}
\bordermatrix{ & I & iX & U \cr I & \Delta_{00} & \Delta_{01} & \frac{\Delta_{00}+\Delta_{01} + \Delta_{10} + \Delta_{11}}{4} \cr 
iX & \Delta_{10} & \Delta_{11} & \frac{\Delta_{00}+\Delta_{01} + \Delta_{10} + \Delta_{11}}{4} \cr 
U & \frac{\Delta_{00}+\Delta_{01} + \Delta_{10} + \Delta_{11}}{4} & \frac{\Delta_{00}+\Delta_{01} + \Delta_{10} + \Delta_{11}}{4} & \frac{\Delta_{00}+\Delta_{01} + \Delta_{10} + \Delta_{11}}{4}
}. 
\end{equation}
\end{enumerate}
The resulting bimatrix games are basically classical, they can be solved with the methods of searching for equilibria known from classical game theory, but in order to play them, players must be equipped with tools that allow them to operate on quantum entangled states. The game (\ref{typeI}) can be identified with the classical game as the third row or column are mixed strategies, where players play the first two strategies $I$ and $iX$ with equal probabilities. Bimatrices (\ref{typeII}) and (\ref{typeIII}) are essentially new games that are not equivalent to the original game. Using the quantum method of manipulating the payoffs of the classical game, players gained access to extended games that allow finding new equilibria significantly more favorable than the equilibria of the original game. In the next section, we show how important this can be to solve the original game.
\section{Nash equilibria in isomorphism-resistant extensions}
In this section, we will demonstrate how the operators determined in the previous section affect Nash equilibria.

Let us now consider a general example of the Prisoner's Dilemma. In this case the payoffs of the bimatrix (\ref{wbimacierz}) have the form
\begin{equation}\label{PD}
\Delta_{00}=(R,R),\hspace{2mm}
\Delta_{01}=(S,T),\hspace{2mm}
\Delta_{10}=(T,S),\hspace{2mm}
\Delta_{11}=(P,P),
\end{equation}
where: $T>R>P>S$ and $2R> T+S$.
Let us consider (\ref{typeII}). Then, taking into account (\ref{PD}), the resulting payoff bimatrix is 
\begin{equation}\label{qq3PD}
\bordermatrix{& I & iX & U \cr I & (R,R) & (S,T) & \left(\frac{T+P}{2},\frac{S+P}{2}\right) \cr
iX & (T,S) & (P,P) & \left(\frac{R+S}{2},\frac{R+T}{2}\right) \cr
U & \left(\frac{S+P}{2},\frac{T+P}{2}\right) & \left(\frac{R+T}{2},\frac{R+S}{2}\right) & \left(\frac{R+S+T+P}{4},\frac{R+S+T+P}{4}\right)}.
\end{equation}
The game has a mixed-strategy Nash equilibrium $((\frac{1}{4},\frac{1}{4},\frac{1}{2}), (\frac{1}{4},\frac{1}{4},\frac{1}{2}))$. Indeed, if one of the players uses the equilibrium strategy, the payoff of the second is constant and equal to $\frac{R+S+T+P}{4}$. We will prove that this is the unique equilibrium in the game. 

Let us first observe that there is no Nash equilibrium in pure strategies. Now, we will show that there is no Nash equilibrium in which the support of the equilibrium strategy consists of two elements. Since the game is symmetric, without loss of generality, let us consider possible Nash equilibrium strategies of the first player. Suppose that a mixed equilibrium strategy of the first player is of the form $(a_{1}, a_{2}, 0)$, $a_{1} + a_{2} =  1$, $a_{1}, a_{2} >0$. In this scenario, the best response of player 2 is to play the strategy of the form $(0, b_{2}, b_{3})$, $b_{2} + b_{3} = 1$, $b_{2}, b_{3} >0$ as
\begin{equation}
u_{2}((a_{1}, a_{2}, 0), I) = a_{1}R + a_{2}S < a_{1}T + a_{2}P = u_{2}((a_{1}, a_{2}, 0), iX).
\end{equation}
However, if player 2's equilibrium strategy took the form $(0, b_{2}, b_{3})$, the second strategy of player 1 would not be part of the support of the best response to $(0, b_{2}, b_{3})$ due to the following inequality:
\begin{equation}
u_{1}(iX, (0, b_{2}, b_{3})) = b_{2}P + b_{3}\frac{R+S}{2} < b_{2}\frac{R+T}{2} + b_{3}\frac{R+S+T+P}{4} = u_{1}(U, (0, b_{2}, b_{3})).
\end{equation}
As a result, a mixed strategy in the form of $(a_{1}, a_{2}, 0)$ cannot be a Nash equilibrium strategy. Similarly, we can show that any other mixed strategy with two-element support cannot be an equilibrium strategy. The Nash equilibrium of the game (\ref{qq3PD}) appears to be a proper quantum solution for the classical Prisoner's Dilemma (\ref{pdkonkret}). Players in the quantum game, equipped with a pair of entangled qubits, by properly mixing the available strategies, can achieve equilibrium $\frac{R+S+T+P}{4}$, which is the expected value of all four payoffs of the classical game with the same probability. This is the only linear combination of payoffs, that is invariant under transpositions of rows and columns of the original game. 
\section{Conclusions}
Variations in the EWL scheme as a game with a finite number of strategies are widely used in the literature on quantum games. Such simplification still satisfies the fundamental requirement imposed on the framework of a quantum game, namely that the output game constitutes a generalization of a classical game. At the same time, a game with a finite number of unitary strategies is more straightforward to examine. This approach allows for a natural transfer of all game theory concepts applied to classical games to quantum game theory. However, not all unitary operations appear to be suitable from the point of view of the desired properties of game extensions. 

In our paper, we have shown that extending a bimatrix game by adding a unitary strategy can lead to multiple different extensions of the same game theory problem. Each of these extensions can imply a completely different rational outcome of the game. In the majority of cases, using any arbitrary unitary strategy generates different payoffs depending on how we number the players' strategies in the classical game. 

We identified finite classes of operators that, for isomorphic classical games, also generate an extension that is identical up to isomorphism. Using such an operator will always yield the same game theory problem, regardless of the order of strategies in the classical game. In this case, we can be certain that the result of the extended game determined by a particular solution concept (eg Nash Equilibrium) is always the same regardless of the
form of the initial game.

One of the practical applications demonstrated in the paper is the extension of the classical Prisoner’s Dilemma game to a quantum game. This quantum version of the game achieves a unique Nash equilibrium that is closer to Pareto-optimal solutions than the original classical game, illustrating a practical benefit of quantum game theory in solving classical game theory dilemmas.


\bibliographystyle{qipstyle}
\bibliography{references}

\begin{thebibliography}{10}
\providecommand{\url}[1]{\texttt{#1}}
\providecommand{\urlprefix}{URL }
\providecommand{\doi}[1]{https://doi.org/#1}

\bibitem{meyer_quantum_1999}
Meyer, D.A.: Quantum {Strategies}. Physical Review Letters  \textbf{82}(5),
  1052--1055 (Feb 1999). \doi{10.1103/PhysRevLett.82.1052},
  \url{https://link.aps.org/doi/10.1103/PhysRevLett.82.1052}, publisher:
  American Physical Society

\bibitem{eisert_quantum_1999}
Eisert, J., Wilkens, M., Lewenstein, M.: Quantum {Games} and {Quantum}
  {Strategies}. Physical Review Letters  \textbf{83}(15),  3077--3080 (Oct
  1999). \doi{10.1103/PhysRevLett.83.3077},
  \url{https://link.aps.org/doi/10.1103/PhysRevLett.83.3077}, publisher:
  American Physical Society

\bibitem{flitney_nash_2007}
Flitney, A.P., Hollenberg, L.C.L.: Nash equilibria in quantum games with
  generalized two-parameter strategies. Physics Letters A  \textbf{363}(5),
  381--388 (Apr 2007). \doi{10.1016/j.physleta.2006.11.044},
  \url{http://www.sciencedirect.com/science/article/pii/S0375960106018184}

\bibitem{elgazzar_quantum_2020}
Elgazzar, A.S.: Quantum prisoner’s dilemma in a restricted one-parameter
  strategic space. Applied Mathematics and Computation  \textbf{370},  124927
  (Apr 2020). \doi{10.1016/j.amc.2019.124927},
  \url{https://linkinghub.elsevier.com/retrieve/pii/S0096300319309191}

\bibitem{frackiewicz_strong_2016}
Frąckiewicz, P.: Strong {Isomorphism} in {Eisert}-{Wilkens}-{Lewenstein}
  {Type} {Quantum} {Games} (Aug 2016).
  \doi{https://doi.org/10.1155/2016/4180864},
  \url{https://www.hindawi.com/journals/amp/2016/4180864/}, iSSN: 1687-9120
  Pages: e4180864 Publisher: Hindawi Volume: 2016

\bibitem{frackiewicz_nash_2022}
Frąckiewicz, P., Szopa, M., Makowski, M., Piotrowski, E.: Nash {Equilibria} of
  {Quantum} {Games} in the {Special} {Two}-{Parameter} {Strategy} {Space}.
  Applied Sciences  \textbf{12}(22),  11530 (Jan 2022).
  \doi{10.3390/app122211530}, \url{https://www.mdpi.com/2076-3417/12/22/11530},
  number: 22 Publisher: Multidisciplinary Digital Publishing Institute

\bibitem{frackiewicz_quantum_2017}
Frąckiewicz, P., Pykacz, J.: Quantum {Games} with {Strategies} {Induced} by
  {Basis} {Change} {Rules}. International Journal of Theoretical Physics
  \textbf{56}(12),  4017--4028 (Dec 2017). \doi{10.1007/s10773-017-3423-6},
  \url{https://doi.org/10.1007/s10773-017-3423-6}

\bibitem{ozdemir_samaritans_2003}
Ozdemir, S.K., Shimamura, J., Morikoshi, F., Imoto, N.: Samaritan's {Dilemma}:
  {Classical} and quantum strategies in {Welfare} {Game} (Nov 2003),
  \url{http://arxiv.org/abs/quant-ph/0311074}, issue: arXiv:quant-ph/0311074
  arXiv:quant-ph/0311074

\bibitem{pawela_quantum_2013}
Pawela, L., Sładkowski, J.: Quantum {Prisoner}'s {Dilemma} game on hypergraph
  networks. Physica A: Statistical Mechanics and its Applications
  \textbf{392}(4),  910--917 (Feb 2013). \doi{10.1016/j.physa.2012.10.034},
  \url{http://arxiv.org/abs/1202.5934}, arXiv:1202.5934 [physics,
  physics:quant-ph]

\bibitem{landsburg_nash_2011}
Landsburg, S.: Nash equilibria in quantum games. Proceedings of the American
  Mathematical Society  \textbf{139}(12),  4423--4434 (2011).
  \doi{10.1090/S0002-9939-2011-10838-4},
  \url{https://www.ams.org/proc/2011-139-12/S0002-9939-2011-10838-4/}

\bibitem{szopa_efficiency_2021}
Szopa, M.: Efficiency of {Classical} and {Quantum} {Games} {Equilibria}.
  Entropy  \textbf{23}(5), ~506 (May 2021). \doi{10.3390/e23050506},
  \url{https://www.mdpi.com/1099-4300/23/5/506}, number: 5 Publisher:
  Multidisciplinary Digital Publishing Institute

\bibitem{pawela_quantum_2016}
Pawela, L.: Quantum games on evolving random networks. Physica A: Statistical
  Mechanics and its Applications  \textbf{458},  179--188 (Sep 2016).
  \doi{10.1016/j.physa.2016.04.022}, \url{http://arxiv.org/abs/1512.09104},
  arXiv:1512.09104 [physics, physics:quant-ph]

\bibitem{yong_entanglement_2016}
Yong, X., {Hong-liang Sun}, {Juan Li}: Entanglement plays an important role in
  evolutionary generalized prisoner's dilemma game on small-world networks. In:
  2016 {IEEE} {Advanced} {Information} {Management}, {Communicates},
  {Electronic} and {Automation} {Control} {Conference} ({IMCEC}). pp. 319--324.
  IEEE, Xi'an, China (Oct 2016). \doi{10.1109/IMCEC.2016.7867225},
  \url{http://ieeexplore.ieee.org/document/7867225/}

\bibitem{li_entanglement_2015}
Li, A., Yong, X.: Entanglement {Guarantees} {Emergence} of {Cooperation} in
  {Quantum} {Prisoner}'s {Dilemma} {Games} on {Networks}. Scientific Reports
  \textbf{4}(1), ~6286 (May 2015). \doi{10.1038/srep06286},
  \url{http://www.nature.com/articles/srep06286}, number: 1

\bibitem{deng_novel_2016}
Deng, X., Zhang, Q., Deng, Y., Wang, Z.: A novel framework of classical and
  quantum prisoner’s dilemma games on coupled networks. Scientific Reports
  \textbf{6}(1),  23024 (Sep 2016). \doi{10.1038/srep23024},
  \url{http://www.nature.com/articles/srep23024}, number: 1

\bibitem{van_enk_classical_2002}
van Enk, S.J., Pike, R.: Classical {Rules} in {Quantum} {Games}. Physical
  Review A  \textbf{66}(2),  024306 (Aug 2002).
  \doi{10.1103/PhysRevA.66.024306},
  \url{http://arxiv.org/abs/quant-ph/0203133}, number: 2 arXiv:quant-ph/0203133

\bibitem{maschler_game_2020}
Maschler, M., Solan, E., Zamir, S.: Game theory. Cambridge University Press,
  Cambridge, UK, (2020), oCLC: 1180191769

\bibitem{gabarro_complexity_2007}
Gabarró, J., García, A., Serna, M.: On the {Complexity} of {Game}
  {Isomorphism}. In: Kučera, L., Kučera, A. (eds.) Mathematical {Foundations}
  of {Computer} {Science} 2007. pp. 559--571. Lecture {Notes} in {Computer}
  {Science}, Springer, Berlin, Heidelberg (2007).
  \doi{10.1007/978-3-540-74456-650}

\bibitem{nash_non-cooperative_1951}
Nash, J.: Non-{Cooperative} {Games}. Annals of Mathematics  \textbf{54}(2),
  286--295 (1951). \doi{10.2307/1969529},
  \url{https://www.jstor.org/stable/1969529}, publisher: Annals of Mathematics

\bibitem{peleg_canonical_1999}
Peleg, B., Rosenmüller, J., Sudhölter, P.: The canonical extensive form of a
  game form: symmetries. In: Alkan, A., Aliprantis, C.D., Yannelis, N.C. (eds.)
  Current {Trends} in {Economics}: {Theory} and {Applications}, pp. 367--387.
  Studies in {Economic} {Theory}, Springer, Berlin, Heidelberg (1999).
  \doi{10.1007/978-3-662-03750-8-22},
  \url{https://doi.org/10.1007/978-3-662-03750-8-22}

\bibitem{sudholter_canonical_2000}
Sudhölter, P., Rosenmüller, J., Peleg, B.: The canonical extensive form of a
  game form: {Part} {II}. {Representation}. Journal of Mathematical Economics
  \textbf{33}(3),  299--338 (Apr 2000). \doi{10.1016/S0304-4068(99)00019-1},
  \url{https://www.sciencedirect.com/science/article/pii/S0304406899000191}

\end{thebibliography}

\end{document}